# Estimating the duration effects in structural responses by a new energy-cycle based parameter

Mohammadreza Mashayekhi[*1], Mojtaba Harati[2a] and Homayoon E. Estekanchi[1b]

[1]*Department of Civil Engineering, Sharif University of Technology, Tehran, Iran*
[2]*Department of Civil Engineering, University of Science and Culture, Rasht, Iran*

**Abstract.** Strong-motion duration and number of cycles of motion are the well-known parameters with which earthquake engineers currently use to take into account the influence of motion duration for the dynamic analysis of structures. In addition to these prevalent means, using the number of nonlinear cycles—which a structure experiences during the motion—is also one another method to find the influence of the motion duration in the response of structures. In this study, a new approach is proposed to determine the number of nonlinear cycles. Moreover, a parameter based on both hysteretic energy and number of nonlinear cycles is likewise suggested to precisely reflect the shaking characteristics of the motion. The proposed parameter is expected to demonstrate a high correlation with the response of structures even for the structures having deteriorative and degraded behaviour during the motion. This matter is examined and confirmed with the application of this parameter to predict the influence of motion duration in the response of several degrading RC frames. It is revealed that the proposed parameter is able to well predict the influence of motion duration and can be used as a criterion to select ground motions for dynamic analysis.

## 1. Introduction

In order to characterize the main factors associated with an arbitrary earthquake motion, not only the intensity and the frequency content of the motion should be considered but also motion duration may be needed to be investigated and incorporated as well (Raghunandan and Liel 2013, Du and Wang 2013). Most contemporary seismic codes provide special criteria based on a pair of M and R, the magnitude and the source to site distance, to select a suite of ground motions as seismic inputs for dynamic time history analysis; however, only a number of rough suggestions are made to consider the motion duration in such a procedure. While the duration of the earthquakes essentially plays an important role in the dynamic analysis of the structures in which degrading behaviors are expected to be encountered during the motion (Hancock and Bommer 2007, Haselton et al. 2011, Khaloo et al. 2016), the approaches taken to reflect its incorporation are somewhat inadequate compared to its immense influence on the structural response (Samanta and Pandey 2018). There are many methods in order to consider the influence of motion duration, evaluating the possible effects of the strong-motion duration on the overall seismic performance of the buildings (see: Raghunandan and Liel 2013, Hancock and Bommer 2007). The strong-motion duration is usually predicted and measured in such that one part of the motion that is thought to represent the strong shaking of an earthquake is extracted, and therefore, the time interval of this part of the motion is typically taken as the duration of the ground motion being investigated (Bradley 2011).

---

∗Corresponding author, Research Associate, Ph.D., E-mail: mmashayekhi67@gmail.com
[a] Lecturer, E-mail: moj.harati@gmail.com
[b] Professor, E-mail: stkanchi@sharif.edu

Moreover, counting the number of cycles of motion is another method to consider motion duration (Mashayekhi and Estekanchi 2012). Several authors pointed out that the number of cycles of motion is a more robust parameter than the strong-motion duration to investigate the possible influence of the duration of earthquakes (Hancock and Bommer 2005). Several methodologies are proposed to measure the number of cycles of a specific motion. Predominant differences between these proposed methods are arisen from a number of sources, including the way they employ to count the cycles (Bommer and Martinez-Periera 1999, Hancock and Bommer 2005). For instance, several approaches count cycles versus their amplitudes while the others count cycles versus their ranges.

When an SDOF system is subjected to a severe ground motion, the input energy of the motion is dissipated through a nonlinear action, where this dissipation is accomplished within several nonlinear cycles. The number of nonlinear cycles is predominantly dependent on the duration of the motion (Bradley 2011). For example, hysteretic energy of the SDOF systems subjected to a pulse-like motion dissipates through roughly one cycle. On the other hand, the associated hysteretic energy of SDOF systems exposed to long-duration ground motions is dissipated in several cycles. Therefore, the number of nonlinear cycles of the SDOF systems under earthquake ground motions can be a commendable predictor of motion duration. In addition to depending on motion duration, the number of nonlinear cycles primarily is more related to the response of structures compared to other conventional approaches measuring motion duration.

In this study, first, a new definition for obtaining the number of cycles is proposed. Then a parameter is put forward based on hysteretic energy and number of nonlinear cycles of an SDOF system to well predict the influence of motion duration in the seismic response of the buildings. As it is well-known among earthquake engineers, motion duration is a secondary parameter and thus making comparison between structural responses would be meaningless if one only considers this characteristic of the motion regardless of its intensity and frequency content. So, to reduce the influence of intensity and frequency content of the motions in structural responses, spectral matching procedure can be selected to be employed as a suitable approach in this case (Hancock and Bommer 2007).

In order to demonstrate the effectiveness of the proposed parameter, three reinforced concrete (RC) frames are taken to be investigated in this study. A methodology is correspondingly recommended to determine the number of nonlinear cycles of typical RC frames. Finally, the aforementioned proposed parameter is calculated for the considered structures and their correlations with responses are investigated in order to reveal its effectiveness to predict the influence of motion duration.

## 2. Reference ground motions

Nonlinear dynamic analysis is becoming the most-widely-used procedure for seismic response assessment of structures. The process pertinent to the selection of ground motions as dynamic loading demands a momentous and challenging consideration whenever this kind of analysis is going to be accomplished because it can strongly influence in the response of structures. However, the mentioned influential potential of the selection process is not considered in the current design codes, and the most contemporary seismic codes, such as ASCE7-05 (2006) standard, prescribe a relatively similar procedure for selection of such seismic inputs.

Seismic motions can be generally represented by real, artificial or even response-spectrum-compatible records since a number of important seismological parameters such as magnitude, distance, and local site conditions are considered to reflect the local seismic scenario. This study employs the far-field records for nonlinear dynamic analysis, which is the recommended set of the ground motions by the FEMA P695 (2009). It is worth mentioning that these records are appropriate for structures found at different sites, i.e. the ones with different ground motion hazard functions, site and source conditions. Therefore, a set of twenty-two ground motions have been taken belonging to a bin of rather large magnitudes of 6.5-7.6 as listed in Table 1.

**Table 1, The suite of twenty-two ground motion records utilized as seismic inputs**

| ID No | Earthquake | | Station |
|---|---|---|---|
| | M | Name | |
| 1 | 6.7 | Northridge | Beverly Hills-Mulhol |
| 2 | 6.7 | Northridge | Canyon Country-WLC |
| 3 | 7.1 | Ducze-Turkey | Bolu |
| 4 | 7.1 | Hector Mine | Hector |
| 5 | 6.5 | Imperial Valley | Delta |
| 6 | 6.5 | Imperial Valley | El Centro Array #11 |
| 7 | 6.9 | Kobe, Japan | Nishi-Akashi |
| 8 | 6.9 | Kobe, Japan | Shin-Osaka |
| 9 | 7.5 | Kocaeli, Turkey | Ducze |
| 10 | 7.5 | Kcaeli, Turkey | Arcelik |
| 11 | 7.3 | Landres | Yermo Fire Station |
| 12 | 7.3 | Landres | Coolwater |
| 13 | 6.9 | Loma Prieta | Capitola |
| 14 | 6.9 | Loma Prieta | Gilory Array #3 |
| 15 | 7.4 | Manjil, Iran | Abbar |
| 16 | 6.5 | Superstition Hills | Elcentro Imp. Co. |
| 17 | 6.5 | Superstition Hills | Poe Road (temp) |
| 18 | 7.0 | Cape Mendocino | Rip Del Overpass |
| 19 | 7.6 | Chi-Chi, Taiwan | CHY101 |
| 20 | 7.6 | Chi-Chi, Taiwan | TCU045 |
| 21 | 6.6 | San Fernando | LA-Hollywood Stor |
| 22 | 6.5 | Friuli, Italy | Tolmezzo |

## 3. Conventional approaches considering the influence of the motion duration

It is commonly accepted among earthquake engineers that strong-motion duration may have significant influence on the seismic response of the buildings (Samanta and Pandey

2018), especially for those structures that experience degrading behaviour during the motion. Regarding the importance of incorporating motion duration into seismic structural assessment, more than 30 definitions are proposed to determine the strong-motion duration (Bommer and Martinez-Periera 1999). The mentioned definitions can be typically categorized into three generic groups, namely the bracketed duration, the uniform duration, and the significant duration as well. However, note that there is no universally accepted definition for determining the strong-motion duration. In this study, the bracketed duration with thresholds of 0.05g and 0.1g, and significant duration with limits ranging from 5% to 95% and 5% to 75% of the total arias intensity of the earthquakes are taken as the metrics for strong-motion duration. Their related abbreviations used hereafter are $tb_{0.05}$, $tb_{0.1}$, $ta_{5-95}$, $ta_{5-75}$, respectively.

Another approach to track the influence of motion duration on the structural responses is to consider the number of cycles of motion. Some authors emphasize that this criterion is more reliable than the strong-motion duration, so a number of approaches for cycle counting are suggested by different authors (e.g., Bommer et al. 2006, Manfredi 2001, Mashayekhi and Estekanchi (2013, 2012)). Similar to the methods pertinent to finding the strong-motion duration, there is no universally acknowledged methodology to count the number of cycles of a motion. The Rainflow counting approach has been adopted hereafter, which is explained in ASTM (1985). The most important point that should be noted about the abovementioned approach is that it counts the cycles that have different ranges. Thus, the number of effective cycles and the damage parameter are used in order to consider cycles with different amplitudes (Malhotra 2002). These parameters are mentioned in Equations 1 and 2, respectively as follows:

$$N_{cy} = \frac{1}{2}\sum_{i=1}^{2tn}\left(\frac{u_i}{u_{max}}\right)^c \quad (1)$$

$$D = C \sum_{i=1}^{2tn} u_i^c \quad (2)$$

Where $u_i$ represents the amplitude of the i-th half cycle, $u_{max}$ stands for the amplitude of largest half cycles, tn is assigned to be as the total number of cycles. C and c are the application dependent damage coefficients, where values of c=2 and C=1 are employed in this research.

**4. A new procedure to count nonlinear cycles considering motion duration**

For a structure subjected to a severe ground motion, a part of the total energy of the earthquake is quite likely to get dissipated through nonlinear mechanisms in which the dissipation process generally occurs in several nonlinear cycles. Whether this process is concentrated in one cycle or in a number of nonlinear cycles strongly depends on shaking characteristics of the input motion since experiencing one cycle is more probable in pulse-like motions while records with longer strong-motion duration are anticipated to produce more number of nonlinear cycles. One approach for calculating the number of nonlinear cycles is to count cycles directly regardless of the associated hysteretic energy dissipated within these cycles. The main disadvantage of this method is that it ignores the influence of hysteretic energy of the cycles in the counting procedure. In order to improve the deficiencies of the aforementioned method, however, only reversal of the nonlinear cycles is proposed to be counted. There is yet another way by which the hysteretic energy of the cycles is taken into account as a part of the counting process (Manfredi 2001). In this method, the hysteretic energy is adjusted and normalized by strength level, the maximum displacement under earthquake motion and yield displacement of the system as expressed in Equation 3.

$$N_{eq} = \frac{E_H}{F_y(x_{max}-x_y)} \qquad (3)$$

It is important to note that this method is developed taking into account an elastic-perfectly plastic system only. But in this paper, a definition is suggested which can be applied to all nonlinear systems, even the ones encounter degrading and deterioration during the motion. In this definition, as a first step, the cycle within which the largest amount of the hysteretic energy would be dissipated is found and called the main cycle henceforth. Next, a number of equivalent cycles, which are the same as the main cycle, are generated in such that they would be capable of dissipating the total hysteretic energy of the system. Therefore, this definition is anticipated to measure the number of dominant nonlinear cycles in seismic responses of the structures. It is so obvious that the number of dominant nonlinear cycles would be rather large for long duration motions, whereas the corresponding value seems to be relatively small for pulse-like motions. This might emphasize on the point that the number of nonlinear cycles may be an appropriate representative of the influence of the strong-motion duration in the seismic response of structures.

$$N_{eq} = \frac{E_H}{E_{H,max}} \qquad (4)$$

In this case, a parameter is also proposed in this study to take account of the influence of strong-motion duration. It is generally believed that motion duration is a secondary parameter, and therefore, it is rather pointless to investigate its influence regardless of the intensity and the frequency content of the motions. The influence of the intensity and frequency content is incorporated within the proposed parameter, enabling us to deal with the strong-motion duration as a primary parameter. It is noteworthy to mention that an SDOF system is more vulnerable to get damaged if more hysteretic energy is demanded from the structure in a few numbers of cycles. So, this issue motivated us to express this parameter as Equation 5.

$$\beta = \frac{E_H}{F_y \delta_y N_{eq}{}^\alpha} \tag{5}$$

Where $E_H$ represents hysteretic energy which is imposed on the structure through motions; $N_{eq}$ is designated as the nonlinear number of cycles; $F_y$ stands for the strength level; $\delta_y$ is the yield displacement, and $\alpha$ is assigned as a constant value which should be obtained to establish a strong correlation between motion duration and responses in the structures.

In order to find an optimum value for α, numerous SDOF structures with degrading behaviour are examined in detail. The characteristics associated with the degrading behaviour of the considered structural systems are fully explained in section 5. These structures have the fundamental structural periods equal to 0.5 seconds, 1 second and 3 seconds to approximately cover nearly all the main periods existed in practice. Furthermore, the strength levels of degrading models are carefully altered to represent various SDOF systems with different nonlinearity behaviour. Consequently, these systems can roughly comprise every structure, from those weakly designed to the ones seemed to be overdesigned. For different values of α, the variation between the responses of these SDOF systems and the β parameter is subsequently explored as shown in Figure 1.

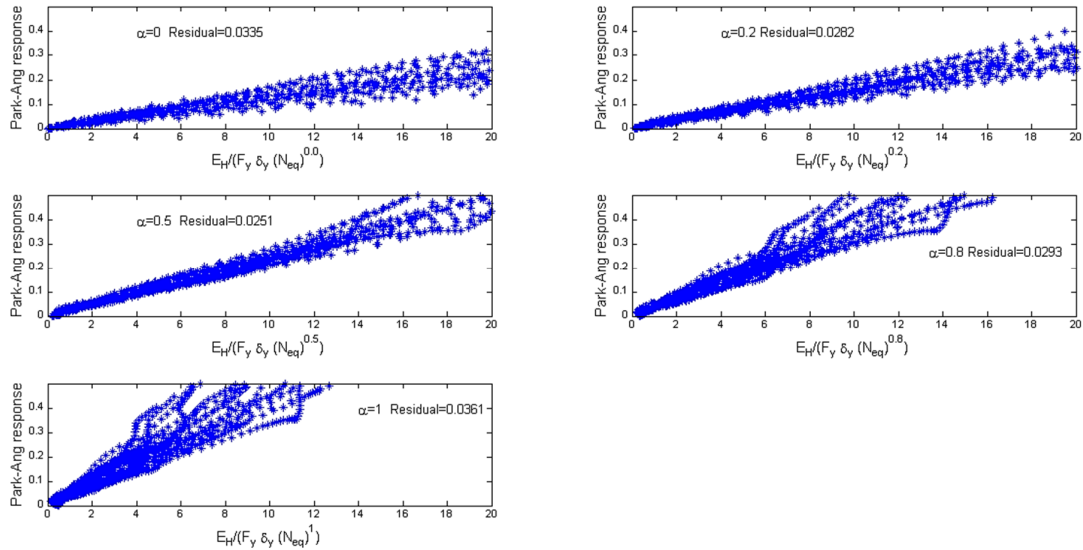

*Fig 1 Variation of response parameter of SDOF systems versus the β parameter for different values of α*

Figure 1 displays the variation of the response of structures against the proposed parameter, the β, with different values of applied α. It is apparently understandable that this parameter can be readily employed to predict the response of structures. It means that in addition to the motion duration, intensity and frequency content are well reflected in this parameter, and therefore, a high correlation between the results has been observed. According to Figure 1, it is obvious that the value of α equal to 0.5 seems to be pretty fine as a result of the minimum value of statistic residual. Thus the β parameter can be rewritten as Equation 6 as follows:

$$\beta = \frac{E_H}{F_y \delta_y N_{eq}^{0.5}} \tag{6}$$

Figure 2 shows the residual of the statistic prediction of response of structure against the proposed parameter for value of α equal to 0.5.

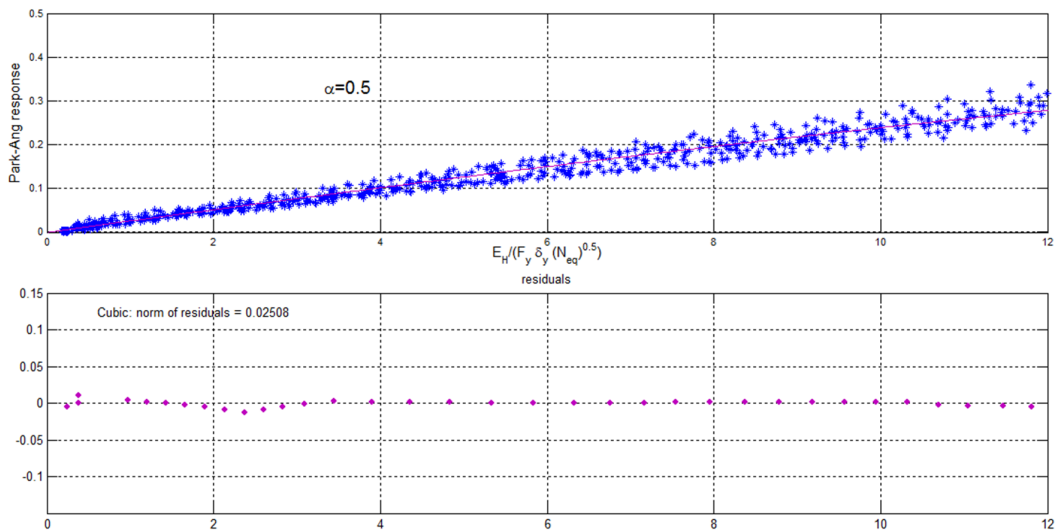

*Fig 2 Residual of prediction of Park-Ang response measure against the proposed parameter*

### 5. Selection of reference models

Since the concrete structures intensely encounter degrading behaviour during the motions, especially in long-duration earthquakes, and degrading structures are more susceptible to motion duration (Hancock and Bommer 2006), several RC frames are employed in this study. For designing the elements of these structures, ACI318-08 is taken as a design code and the related seismic loading is specified based on Iranian National Building Code-Part 6 (2005) for soil type 2 (Similar to the soil type C in NERHP provision). In this case, the lateral resisting frame for these structures is of intermediate type with response modification factor equal to 7. Three low to medium rise frames have been chosen for this study, namely one bay-3 story, one bay-6 story, and one bay-10 story. Configurations of these structures including the height of stories, length of bays and section of beams and columns are illustrated in Figure 2.

Lumped plasticity model is utilized to take the nonlinear behaviour of RC structures into account. As it is well recognized, the nonlinearity of RC structures can be spread over all sections of beams and columns due to the cracking phenomenon. Nonlinear springs at the two ends of the elements are employed to consider the nonlinear behaviour occurring in the structures. Characteristics of the nonlinear springs are described with the peak-oriented model which is early developed by Clough and Johnston (1966) and later modified by Mahin and Stephen (1976). The peak-oriented model is depicted in Figure 3. In this study, the hysteretic model proposed by Ibarra et al. (2005) is used, which includes four deterioration modes: namely, basic strength deterioration, post-capping strength deterioration, unloading stiffness deterioration and accelerated reloading stiffness deterioration as well. This model has been already implemented in Opensees software (2014). The softening branch of Ibarra model for the RC structures is associated with concrete crushing, rebar buckling, and fracture and/or bond failure. In order to assign this model to an element, seven parameters namely, $M_y$, $\theta_y$, $\frac{M_c}{M_y}$, $\theta_{cap,P1}$, $\theta_{pc}$, $\lambda$, c should be determined respectively. The monotonic behaviour of the component model is displayed in Figure 4. In order to quantify the parameters which are required as inputs for the lumped plasticity model, the empirical equations recommended in a technical PEER report by Haselton and Deierlein (2008) are developed and used to relate the design and modelling parameters of beam-column elements. For instance, the proposed equation for predicting post-capping rotation capacity is expressed in Equation 7 as follows:

$$\theta_{pc} = (0.76)(0.031)^{\upsilon}(0.02 + 40\rho_{sh}) \leq 0.10 \qquad (7)$$

Where $\upsilon$ stands for axial load ratio and $\rho_{sh}$ represents transverse steel ratio.

Flexural strength ($M_y$) is predicted in this paper by the equations published by Panagiotakos and Fardis (2001) because it was reported to work properly as recommended by FEMA P695 code as well as the annual PEER report. Other parameters are predicted by the equations published by annual PEER reports by Haselton et al. (2008).

To quantify the response of the considered RC frames, different damage indices can be employed (Xu et al. 2018). Among them, four damage indices are considered and included for this investigation, namely the maximum inter-story drift ratio, Park-Ang damage index (Park and Ang 1985) and two other damage indices that are recommended by Bozorgnia and Bertero (2002). As it is quite common in earthquake engineering, extreme damage indices are not affected by motion duration; on the other hand, damage indices based on hysteretic energy and cumulative damages are more susceptible to motion duration. In the

Park-Ang index, the damage is expressed as a linear combination of the maximum deformation and the effects reflecting accumulating nature of the repeated cyclic loading. This damage index is presented in Equation 8 as follows:

$$DI_{Park-Ang} = \frac{x_{max} - x_y}{x_{mon} - x_y} + \beta \frac{E_H}{F_y x_y} \tag{8}$$

Where $x_{max}$, $x_y$, $x_{mon}$, $E_H$, $F_y$, $\beta$ are the maximum displacement under earthquake loading, yield displacement, maximum displacement under monotonically increasing displacement loading, hysteretic energy, level of strength of the structure and non-dimension coefficient, respectively.

Bozorgnia and Bertero (2002) similarly developed two damage indices for generic inelastic SDOF systems. These damage indices are formulated in Equation 9 and 10 as follows:

$$DI_1 = \left[\frac{(1-\alpha_1)(\mu - \mu_e)}{\mu_{mon} - 1}\right] + \alpha_1 \left(\frac{E_H}{E_{mon}}\right) \tag{9}$$

$$DI_2 = \left[\frac{(1-\alpha_2)(\mu - \mu_e)}{\mu_{mon} - 1}\right] + \alpha_2 \left(\frac{E_H}{E_{mon}}\right)^{0.5} \tag{10}$$

*Where:*

$$\mu = \frac{u_{max}}{u_y} \tag{11}$$

$$\mu_e = \frac{u_{elastic}}{u_y} \tag{12}$$

In the above set of equations, $\mu$ represents the displacement ductility factor, $\mu_e$ is designated as the maximum elastic portion of the deformation, $E_H$ stands for the total dissipated energy and $E_{mon}$ is the capacity of hysteretic energy under monotonic loading.

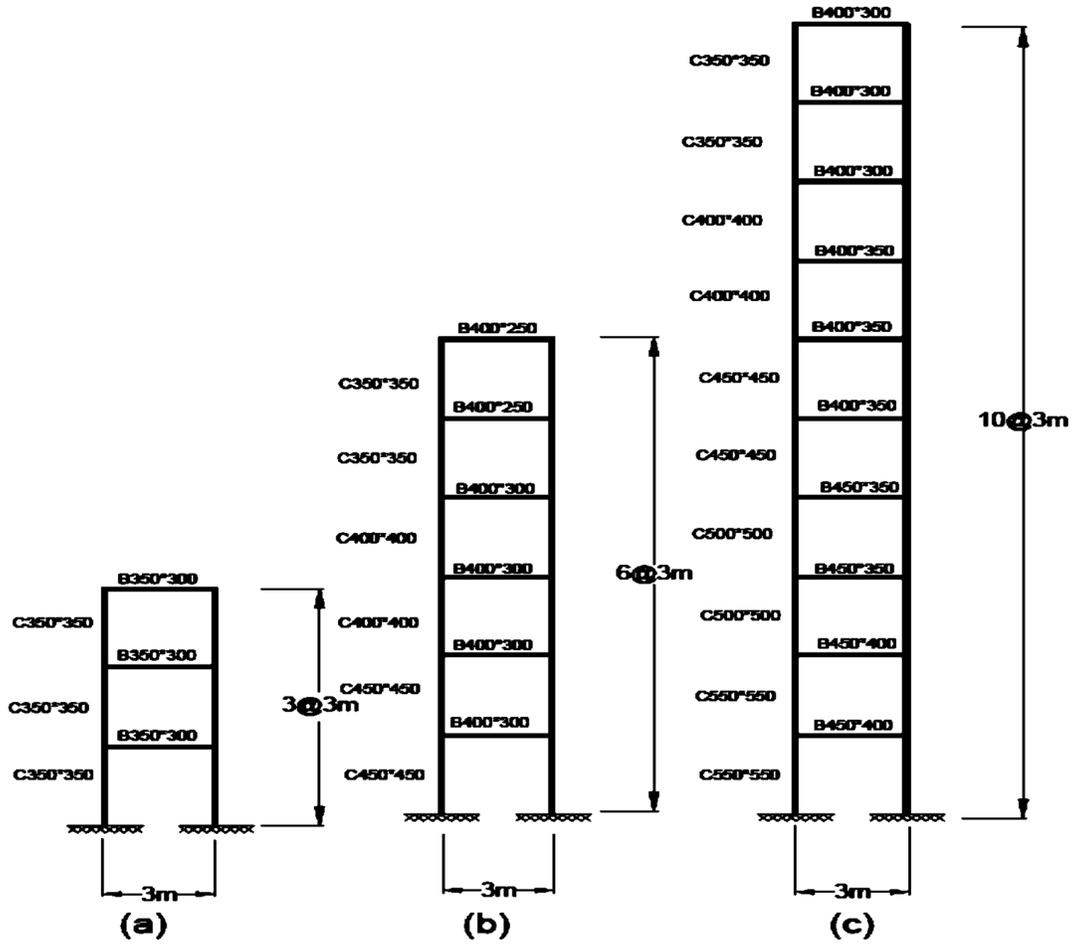

*Fig 3 The considered reference frames: (a) B1S3; (b) B1S6; (c) B1S10*

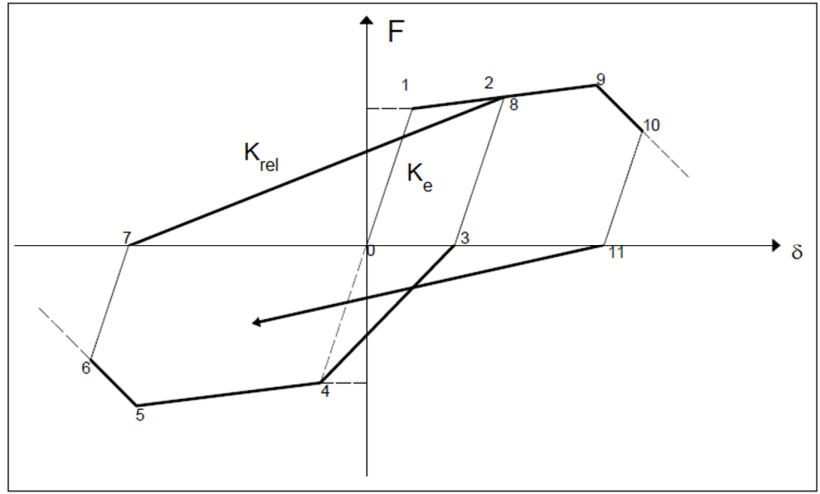

*Fig 4 A basic cyclic backbone for the peak-oriented hysteretic model*

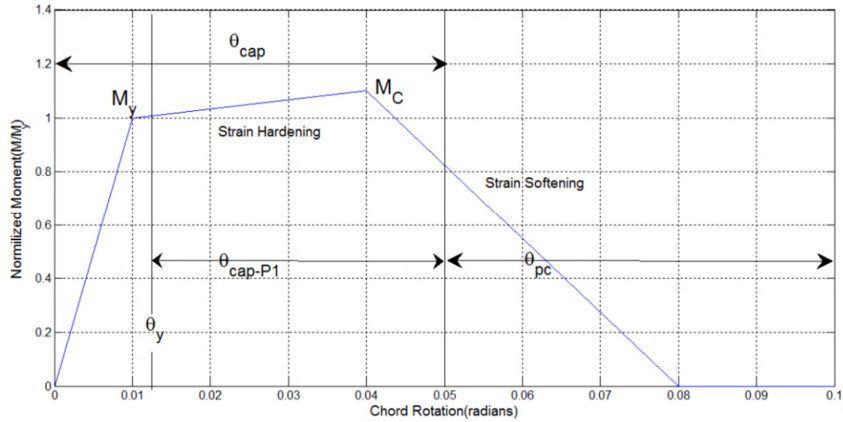

*Fig 5 Monotonic behavior of the component model used to model beam-column elements*

The parameter proposed by the authors, the $\beta$, can be readily computed for any RC structure. In this case, it is worth mentioning that $F_y$ and $\delta_y$ is the constant quantity for a particular RC structure, which can be found through pushover analysis. Since the $\beta$ parameter is usually employed to compare the duration characteristics of different motions, for the sake of simplicity, the abovementioned constant quantities can be therefore removed from Equation 6 when there is only one structure under consideration. The number of nonlinear cycles, $N_{eq}$, of an RC structure can then be determined using Equation 13; moreover, the total dissipated hysteretic energy ($E_H$) can likewise be computed by calculating the sum of all the hysteretic energy dissipated in each spring. So regarding Equation 6 for an RC structure, the $\beta$ parameter can be reliably estimated

using Equation 14 as follows:

$$\beta = \frac{E_H}{\sqrt{N}} \qquad (14)$$

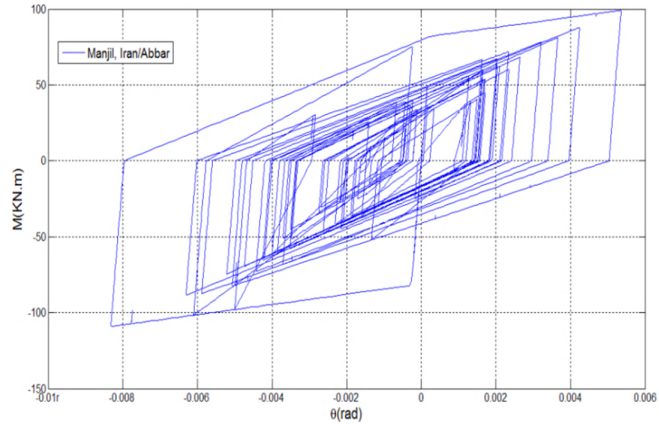

*Fig 6 Moment-rotation response associated with a spring of the beam located at the first story of B1S6 frame subjected to the ground motion of Manjil earthquake of 1990.*

## 7. Results of the regression analysis

In order to diminish the influence of the intensity and the frequency content of motions on the results, the acceleration spectra of the employed ground motions are matched to a reference code spectrum. The unmatched or real acceleration spectra of the selected ground motions are individually depicted in Figure 7; moreover, the acceleration spectra of the matched ground motions are also demonstrated in Figure 8.

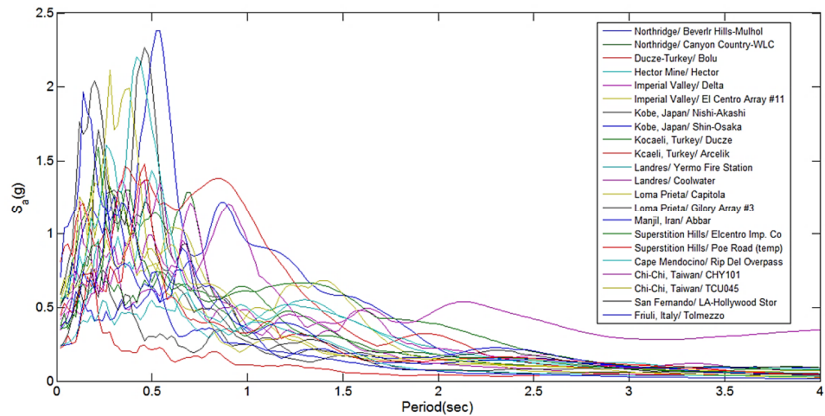

*Fig 7 Acceleration spectra of the employed ground motions*

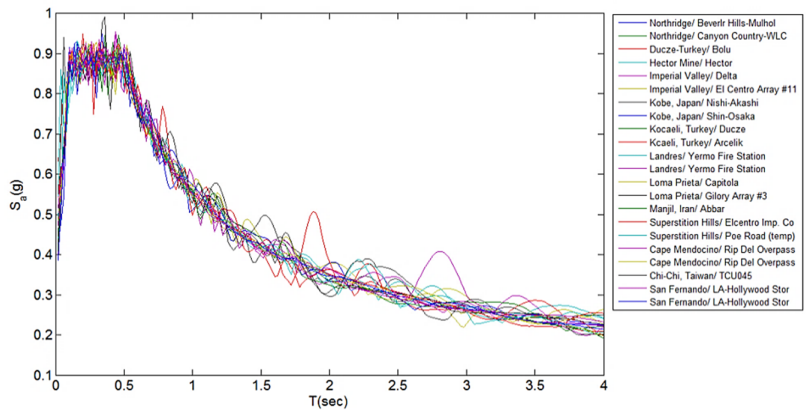

*Fig 8 Acceleration spectra of the matched ground motions*

To investigate the influence of each motion-duration parameter on structural damage measures, their correlations are observed and compared in this section. Since the influence of the intensity and frequency content of the considered ground motions can be ignored as a result of spectral matching procedure, responses are expected to display a positive correlation with the motion-duration parameters. For B1S6 frame, the Park-Ang response measure is plotted against the motion durations obtained from bracketed duration definition with a threshold of 0.05g and the parameter (β) proposed in this study as illustrated in Figures 9 and 10.

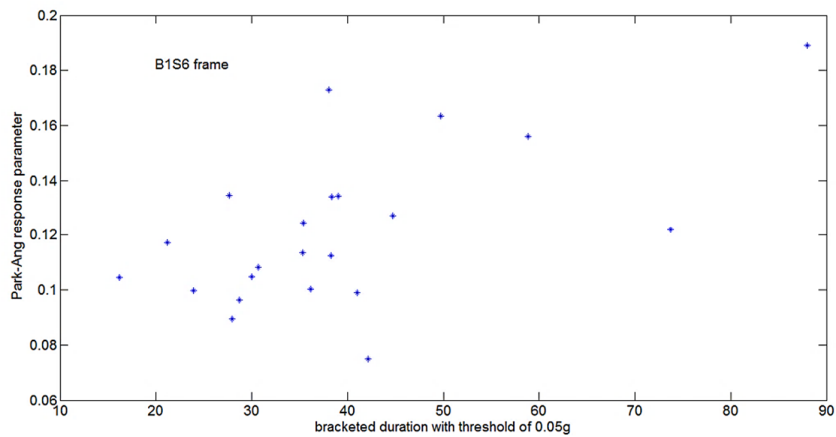

*Fig 9 Variation of response parameter of Park-Ang versus the bracketed duration with a threshold of 0.05g for B1S6 frame.*

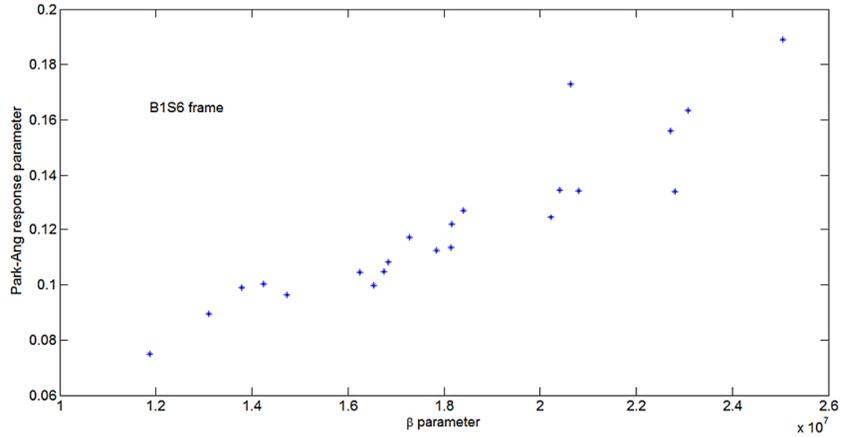

*Fig 10 Variation of response parameter of Park-Ang with the β as a duration measure for B1S 6 frame*

It can be apparently realized from Figure 8 and 9 that the Park-Ang response parameter has a positive correlation with both the bracketed duration and the β parameter as a duration measure; however, its correlation with the β parameter is more remarkable than the bracketed duration. The correlation coefficients between the response measures and the β parameters pertinent to B1S3, B1S6 and B1S10 frames are represented in Table 2.

*Table 2 Correlation coefficients between different response measures and the β parameter*

|  | B1S3 | B1S6 | B1S10 | Average |
|---|---|---|---|---|
| **DR** | 0.27 | 0.10 | 0.67 | 0.35 |
| **Park-Ang** | 0.87 | 0.92 | 0.94 | 0.91 |
| **Bozorgnia-Bertero 1** | 0.68 | 0.82 | 0.86 | 0.79 |
| **Bozorgnia-Bertero2** | 0.65 | 0.86 | 0.90 | 0.80 |

The average values of the correlation coefficients, calculated for all the selected frames with respect to different duration related parameters, is compared with each other as presented in Table 3. Scrutinizing the averages indicated in Table 2 can show us to realize whether or not one response parameter is sensitive to the duration of motion. We can also find which of the duration related parameters proves to be the best predictor of the influence of motion duration on structural responses.

*Table 3 Comparison of the average of the correlation coefficients, considering all the selected structures, versus different duration related parameters*

|  | $tb_{0.05}$ | $tb_{0.1}$ | $ta_{5-75}$ | $ta_{5-95}$ | $N_{cy}$ | D | β |
|---|---|---|---|---|---|---|---|
| **DR** | -0.03 | -0.21 | -0.03 | 0.00 | -0.31 | -0.27 | 0.35 |
| **Park-Ang** | 0.51 | 0.67 | 0.51 | 0.51 | 0.42 | 0.41 | 0.91 |
| **Bozorgnia-Bertero 1** | 0.48 | 0.51 | 0.49 | 0.51 | 0.30 | 0.31 | 0.79 |
| **Bozorgnia-Bertero 2** | 0.45 | 0.47 | 0.47 | 0.49 | 0.27 | 0.29 | 0.80 |

It is revealed from Table 3 that the employed extreme damage index, namely the maximum inter-story drift, is not sensitive to the duration of input motion; however, those damage indices that are based on cumulative damages or the hysteretic energy are strongly affected by duration of the input motions as can be found from the computed averages indicated in Table 3. But it is of worth of mention that the responses, even the maximum inter-story drift, have a high correlation with the β parameter compared to the other motion duration parameters incorporated in this paper. This actually demonstrates the overall effectiveness of the proposed definition and its great potential as a criterion for the record selection procedure.

**8. Conclusion**

1- All current definitions for cycles counting available in the literature have some deficiencies that are presented in the text. In this study, a new procedure for counting the number of nonlinear cycles of an SDOF system is proposed. Besides, this procedure can be accurately applied to systems which encounter degrading behaviour during the motions.

2- A parameter is suggested to independently reflect the influence of strong-motion duration in counting the number of nonlinear cycles of a structure. This parameter, the β, is a combination of hysteretic energy dissipated in the system and the related experienced number of nonlinear cycles during the motions. It can be logically deduced that hysteretic energy and number of cycles have a direct and inverse relationship with structural responses, respectively. So, the hysteretic energy term is in the numerator and the number of cycles is positioned in the denominator of the above-mentioned proposed parameter. Furthermore, the β is optimized and tuned in a way that the best value for the exponent term, belonging to the number of cycles, in the denominator is obtained.

3. A method is also proposed to quantify the number of cycles for an RC frame during the motion. This method is based on a weighted average, which considers the number of cycles of each plastic joint with the associated hysteretic energy dissipated within that joint. By using this approach, earthquake engineers would be able to anticipate the number of nonlinear cycles of a structure for future earthquakes.

4. The proposed parameter—the β—is also calculated for RC frames, and its correlation with the response measures has shown to be high in such cases.

5. The correlation coefficient of the other duration-dependent intensity measures with the different employed damage indices is compared with the one obtained by the parameter proposed in this study, and therefore, it is found that the β parameter can serve as a potential criterion in the record selection procedure.

6. It is re-confirmed that extreme damage index such as the maximum inter-story drift is not susceptible to the duration of input motions when the β parameter is applied to the structures; however, those damage indices that are based on cumulative damages or hysteretic energy are observed to get strongly affected by the duration of motions.

**Acknowledgments**

The authors would like to thank Sharif University of Technology Research Council for their support of this research.

**Nomenclature**

| | |
|---|---|
| C | Application dependent damage coefficients |
| c | Application dependent damage coefficients |
| c | Cyclic deterioration calibration term |
| D | Absolute parameter of number of cycles |
| $DI_1$ | Bozorgnia-Bertero Damage index |
| $DI_2$ | Bozorgnia-Bertero Damage Index |
| $E_h$ | Hysteretic energy |
| $E_{hmax}$ | Maximum hysteretic energy |
| $E_{h,u}$ | Allowable hysteretic energy of the analyzed structure |
| $E_{mon}$ | Capacity of hysteretic energy under monotonic loading |
| $F_y$ | Strength of structure |
| g | Acceleration due to the gravity |
| N | The number of structures which are considered in this study |
| $N_{cy}$ | Relative definition of number of cycles |
| $n_{eq}$ | Equivalent number of cycles |
| $M_y$ | Flexural strength |
| $S_{aC}(T)$ | Target spectrum |
| $S_{aC}(T,t)$ | Acceleration spectrum to be induced at time t |
| $S_{uC}(T,t)$ | Target displacement spectrum at time t |
| T | Free Vibration Period |
| t | Time |
| $ta_{5-95}$ | Significant duration with threshold of 5% to 95% of total arias intensity |
| $ta_{5-75}$ | Significant duration with threshold of 5% to 75% of total arias intensity |
| $tb_{0.05}$ | Bracketed duration with threshold of 0.05g |
| $tb_{0.1}$ | Bracketed duration with threshold of 0.1g |
| tn | Total number of cycles |
| $u_i$ | Amplitude of i-th half cycle |
| $u_{max}$ | Amplitude of largest half cycle |
| $x_{max}$ | Maximum displacement |
| $x_y$ | Yield displacement |
| $\alpha_1$ | Constant coefficient in Bozorgnia-Bertero Damage index |
| $\alpha_2$ | Constant coefficient in Bozorgnia-Bertero Damage index |
| $\beta$ | Duration related parameter |
| $\beta$ | Non-dimension parameter in Park-Ang damage index formula |
| $\delta$ | Comparative parameter |
| $\theta_{cap,P1}$ | Plastic chord rotation from yield to cap |
| $\theta_{cap}$ | Total chord rotation at capping; sum of elastic and plastic deformation (radian) |
| $\theta_{pc}$ | Post capping rotation |
| $\theta_y$ | Yield rotation |
| $\mu_e$ | Maximum elastic portion of deformation |
| $\mu_{mon}$ | Capacity of ductility under monotonic loading |
| $\mu_{\theta u}$ | Ultimate rotation ductility under a monotonic static load |
| $\mu_{\theta m}$ | Maximum rotation ductility |
| $\mu_{\theta y}$ | Yield rotation ductility |
| $\lambda$ | Normalized energy dissipation capacity |
| $\upsilon$ | Axial load ratio |
| $\rho_{sh}$ | Transverse steel ratio. |